\documentclass{pazhastl}
\usepackage{graphicx}
\usepackage{natbib}

\usepackage{amsfonts}
\usepackage{amssymb}

\def\smfigurewocap#1#2#3{
  \begin{minipage}{0.04\linewidth}
    \rotatebox{90}{#3}
  \end{minipage}
  \begin{minipage}{0.9\linewidth}
    \includegraphics[bb=45 185 560 688,width=0.97\linewidth]{#1}
    \centerline{#2}
  \end{minipage}
}

\def\smfigurewocapsmall#1#2#3{
  \begin{minipage}{0.04\linewidth}
    \rotatebox{90}{#3}
  \end{minipage}
  \begin{minipage}{0.9\linewidth}
    \includegraphics[bb=45 185 560 510,width=0.97\linewidth]{#1}
    \centerline{#2}
  \end{minipage}
}

\begin{document}

\journalinfo{2007}{33}{12}{797}[803]

\title{Observations of GRB\,060526 Optical Afterglow with Russian-Turkish
  1.5-m Telescope}

\author{I.~M.~Khamitov\address{1}\email{irekk@tug.tug.tubitak.gov.tr},
      R.~A.~Burenin\address{2},
      I.~F.~Bikmaev\address{3,4},
      N.~A.~Sakhibullin\address{3,4},
      M.~N.~Pavlinsky\address{2},
      R.~A.~Sunyaev\address{2},
      Z.~Aslan\address{1}
      \addresstext{1}{T\"{U}B\.{I}TAK National Observatory, Akdeniz
        Universitesi Yerleskesi, Antalya, Turkey}
      \addresstext{2}{Space Research Institute (IKI), ul. Profsoyuznaya 84/32,
        Moscow, Russia}
      \addresstext{3}{Kazan State University, ul. Kremlevskaya 18, Kazan,
        Russia} 
      \addresstext{4}{Academy of Sciences of Tatarstan, ul.  Baumana, 20,
        Kazan, Russia}
}

\shortauthor{Khamitov \etal}

\shorttitle{RTT150 observations of GRB\,060526 optical afterglow}

\submitted{June 22, 2007}

\begin{abstract}
  
  We present the results of the photometric multicolor observations of GRB
  060526 optical afterglow obtained with Russian-Turkish 1.5-m Telescope
  (RTT150, Mt.\ Bakirlitepe, Turkey). The detailed measurements of afterglow
  light curve, starting from about 5 hours after the GRB and during 5
  consecutive nights were done. In addition, upper limits on the fast
  variability of the afterglow during the first night of observations were
  obtained and the history of afterglow color variations was measured in
  detail. In the time interval from 6 to 16 hours after the burst, there is
  a gradual flux decay, which can be described approximately as a power law
  with an index of $-1.14\pm0.02$. After that the variability on the time
  scale \textit{$\delta t < t$} is observed and the afterglow started to
  decay faster. The color of the afterglow, $V\!\!-\!\!R\approx0.5$, is
  approximately the same during all our observations. The variability is
  detected on time scales up to $\delta t/t\approx 0.0055$ at $\Delta
  F_\nu/F_\nu\approx 0.3$, which violates some constraints on the
  variability of the observed emission from ultrarelativistic jet obtained
  by \cite{ioka}. We suggest to explain this variability by the fact that
  the motion of the emitting shell is no longer ultrarelativistic at this
  time.
  
  
\end{abstract}

\section*{INTRODUCTION}

More than a hundred gamma-ray bursts (GRBs) optical afterglows have been
detected to date, mostly after observations of GRBs by HETE-II \citep{hete2}
and SWIFT \citep{swift} satellites. The photometric observations of GRB
optical afterglows are largely consist of only a few measurements on time
scales from several hours to days. The number of afterglows with the light
curves measured in more detail is not large and each of them is of
considerable interest.

In this paper we present our multicolor photometry of the optical afterglow
from GRB 060526, carried out with Russian-Turkish 1.5-m Telescope (RTT150)
at the T\"UB\.ITAK National Observatory (Mount Bakirlitepe, Turkey). With
RTT150 telescope it was possible to obtain the detailed measurements of
afterglow light curve, starting from about 5 hours after the GRB and during
5 consecutive nights. We obtained more than 200 afterglow brightness
measurements in the first night and more than 20 sensitive measurements in
the subsequent nights.

\section*{Gamma-ray burst 060526}

GRB 060526 was detected by \textit{SWIFT} instruments on May 26, 2006, at
16:28:30 UT. The gamma-ray burst demonstrated a multipeak light curve with a
duration of about 20~s, then a well separated second period of activity
starting after about 200~s was observed. According to its characteristics in
gamma-rays \citep{markwardt}, it was a typical ``long'' burst. The X-ray
and optical telescopes of \textit{SWIFT} satellite were pointed to the burst
about 70 sec after the trigger and found previously unknown X-ray and
optical source \citep{campana}.

The ground-based observations of the optical transient were started as early
as 36 sec after the beginning of gamma-ray burst using an automatic 40-cm
telescope \citep{french} and were continued using larger telescopes
\citep{khamitov,dai,covino,lin,greco,thoene,cobb,rumyantsev,
  kann1,baliyan,terra,kann2,sharapov,rumyantsev2}. The redshift of the GRB
source, $z=3.21$, was measured later using 6.5-m Magellan telescope
\citep{berger}.

Considerable part of all photometric measurements of the light curve was
obtained with the Russian-Turkish 1.5-m Telescope. The results of these
observations are presented in this paper.

\begin{figure}
  \centering
  \includegraphics[width=0.9\columnwidth,]{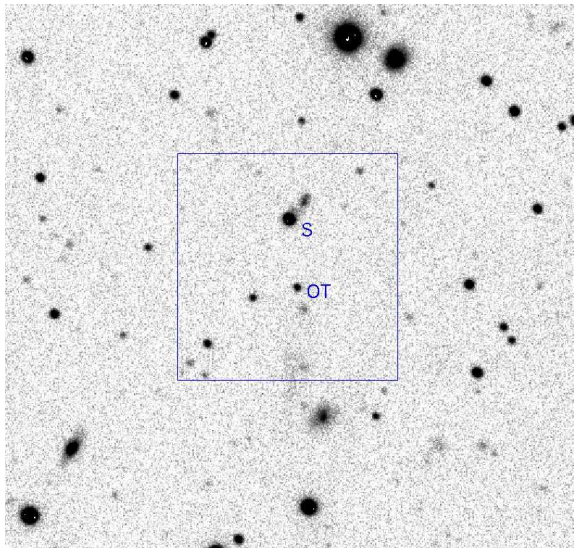}
  \caption{The \textit{$R_c$} image of the GRB 060526 field at the beginning
    of observations. The optical transient (OT) and our reference star (S),
    and the boundary of sub-image used for fast photometry measurements are
    shown.}
  \label{fchart}
\end{figure}

\section*{OBSERVATIONS}

The observational time on the telescope RTT150 was scheduled in frames
target of opportunity observations (TOO) program, according to the agreement
between all participants of the project. Observations of the afterglow field
were performed using CCD-photometer, based on thermoelectrically cooled
Andor $2048\times2048$ CCD. The observations were made with $2\times2$
binning, which gives 0.48\arcsec\ per bin angular scale.

Because of insufficiently good weather conditions on the night of May 26,
the telescope could not be pointed at the afterglow immediately after the
Bacodine alert was received. The telescope was pointed at the GRB field only
5.44 h later. The entire field was immediately imaged in the $B\,V R_c$
bands. The image obtained at this time is shown in Fig.~\ref{fchart}. This
figure also shows the star used in the observations as a reference and the
optical transient as it was during the first hours after the GRB.

Once the optical transient was detected in all images, a smaller field
around it with size $200\times200$ pixels ($\approx 1.5\arcmin \times
1.5\arcmin$, Fig.~\ref{fchart}) was chosen and a few sets of the 30 short
30~s exposures were made. In this mode CCD readout time was 10~s. In order
to measure the colors of the afterglow, several 300~s exposures in
\textit{B} and \textit{V} bands were made between the sets.

\begin{figure}
  \centering
  \smfigurewocap{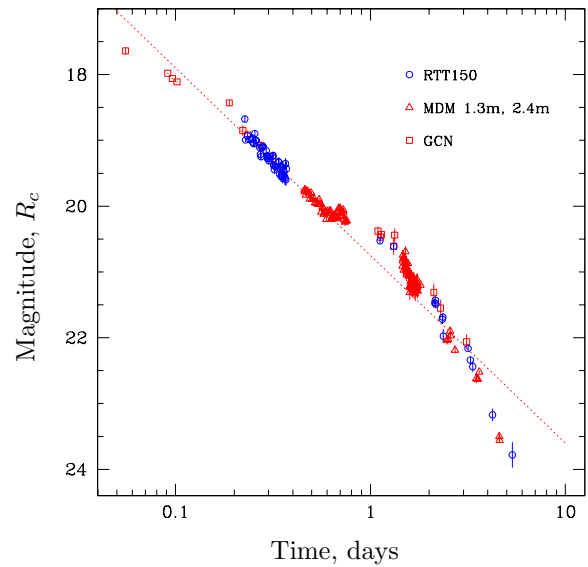}{Time, days}{Magnitude, $R_c$}
  \caption{The lightcurve of the afterglow in \textit{$R_c$} band. The
    circles show the measurements obtained with RTT150 telescope, triangles
    --- data from 1.3-m and 2.4-m \textit{MDM} telescopes \citep{dai},
    squares --- data taken from various GCN Circulars
    \citep{lin,rumyantsev2,terra,kann1}.}
  \label{lc}
\end{figure}

During the next night the field was imaged in \textit{$BVR_c$} bands.  In
subsequent nights from 28 to 31 May, the field was observed in
\textit{$VR_c$} bands. Observations were made only in \textit{$R_c$} band in
the nights of $1^{st}$ and $2^{nd}$ June. Observations were continued until
the afterglow faded below the detection limit of telescope in a few hours of
the exposure. In order to obtain photometrical calibrations, the
observations of \cite{landolt} photometric standards were carried out every
night. The data were processed using standard IRAF
software\footnote{http://iraf.noao.edu}. 

For our photometric measurements the sizes of the apertures were selected to
achieve best signal-to-noise ratio. The magnitudes were obtained based on
the photometric solutions for \cite{landolt} standard stars. For the
reference star, shown in Fig.~\ref{fchart}, the following magnitudes were
obtained: \textit{B}=17.67, \textit{V}=16.90, \textit{R$_c$}=16.48.

\section*{ RESULTS}

In Fig.~\ref{lc} the light curve of the afterglow for the entire period of
observation is shown. The data in time interval from 5 to 14 hours after the
burst are well fitted by a power law $F_\nu \propto t^\alpha$ with the index
$\alpha = -1.14 \pm 0.02$, in agreement with the results of \cite{dai}. This
dependence is shown in Fig.~\ref{lc} by dotted line. At $t-t_0<5$~h, the
decline of the light curve is flatter; at $t-t_0>14$~h, a re-brightening,
some variability, and then faster decay of the optical afterglow flux are
observed. Closer to the end of our observations, the afterglow decays
rapidly. For $t-t_0>3$~days, the fading corresponds to the power law with
index $\alpha = -3.39 \pm 0.10$, also in agreement with \cite{dai}. We note
that the RTT150 observations comprise about a half of the useful
measurements of the afterglow magnitude for this burst.

\begin{figure}
  \centering
  \smfigurewocap{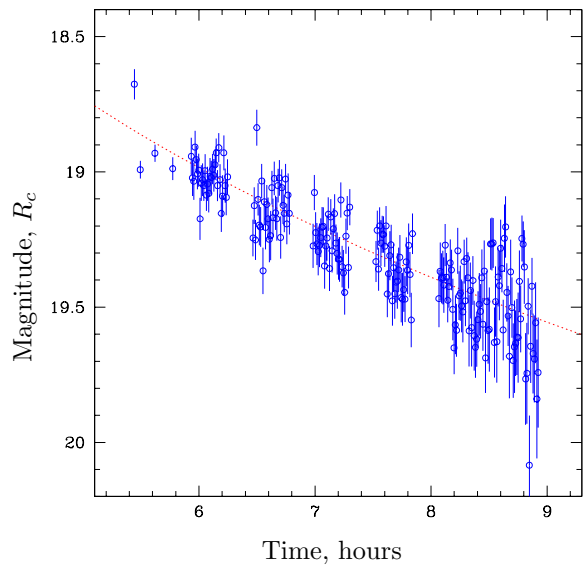}{Time, hours}{Magnitude, $R_c$}
  \caption{The lightcurve of GRB 060526 afterglow in \textit{$R_c$} band in
    the first night of our observations with 40~s time resolution. The power
    law flux decay with index $-1.30$ is shown with dashed line.}
  \label{lc1d}
\end{figure}

During the first night of our observations, more than 200 flux measurements
were made in \textit{$BVR_c$} bands. The results of the light curve
measurements for the first night in $R_c$ band are presented in
Table~\ref{tab:mags1dfast} and the corresponding light curve is shown in
Fig.~\ref{lc1d}. Taking into account the smooth flux decay, the statistical
dispersion of measurement is $0.09^{m}$, which increases a little during the
last hour of observations due to the increasing sky background at dawn. Two
points close to 5.4 h and 6.4 h, respectively, deviate from the general
series as if the afterglow brightness increased at this moment by 20--30\%.
It is possible that this is due to some systematic errors. However, we could
not determine their origin.

\begin{table}
  \renewcommand{\tabcolsep}{0.8cm}
  \centering
  \caption{The photometric measurements in $R_c$ band made
    in the first night of observations with high time
    resolution.\label{tab:mags1dfast}} 
\smallskip
  \begin{tabular}{ccc}
  \hline
  \hline
  Time, h & $m_{R_c}$ & error\\
  \hline
5.4430 & 18.676 &  0.056\\
5.4941 & 18.992 &  0.033\\
5.6201 & 18.931 &  0.032\\
5.7739 & 18.988 &  0.042\\
5.9378 & 18.943 &  0.068 \\
5.9485 & 19.023 &  0.066 \\
5.9594 & 19.035 &  0.068 \\
5.9702 & 18.909 &  0.059 \\
5.9809 & 18.955 &  0.050 \\
5.9917 & 19.008 &  0.056 \\
\dots & \dots & \dots\\
  \hline
  \end{tabular}
  \medskip

  \begin{minipage}{0.9\linewidth}
    \footnotesize \emph{Note}: --- The complete version of this table
    is available in electronic version of the journal, and also at:
    http://hea.iki.rssi.ru/grb/060526/indexeng.html
  \end{minipage}
\end{table}

The agreement between the dispersion of magnitudes measurements shown in
Fig.~\ref{lc1d} and the statistical errors, calculated from the data, is
sufficiently good. That means that there are essentially no unaccounted
systematic errors in our measurements. In addition, it suggests that the
afterglow flux is not variable on time scale $\approx40$~s. The value of the
scatter gives a conservative upper limit on the fast variability of the
afterglow on that time scale, $<10\%$. This agrees well with the
observations GRB 030329 afterglow, when even more stringent constraints were
obtained for the fast variability at the early afterglow phase
\citep{burenin03}.

\begin{figure}
  \centering 
  \smfigurewocap{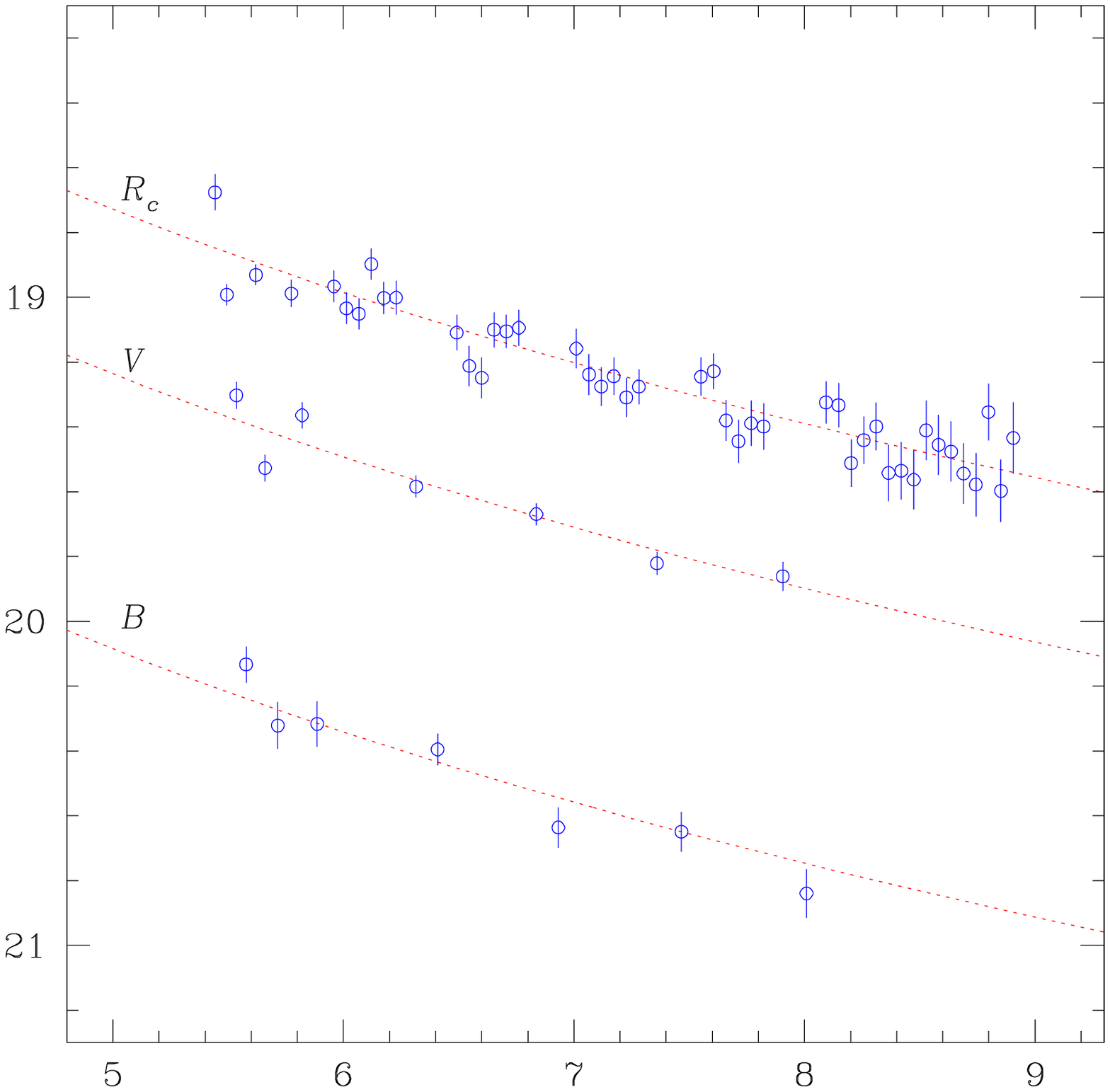}{Time, hours}{Magnitude}
  \caption{The lightcurve of GRB 060526 afterglow in $BVR_{c}$ bands in
    first night of observations. The power law with an index $-1.30$
    obtained using $R_c$ lightcurve is shown by the dotted line. The
    afterglow decay in $V$ and $B$ bands is also in agreement with this
    power law.}
  \label{lc1dbvr}
\end{figure}

The light curves of the afterglow in $BVR_{c}$ bands are shown in
Fig.~\ref{lc1dbvr}. For the $R_{c}$ band, the values at each point were
calculated using five combined images to decrease possible systematic
errors. The fitting of these light curves with power laws gives the indices
$-1.61 \pm 0.18$, $-1.37 \pm 0.11$, and $-1.30 \pm 0.08$ in $B$, $V$, and
$R_c$ bands, respectively. These values are in agreement with each other
within the errors of measurements. However, they are slightly different
from the index $-1.14 \pm 0.02$ given above based on the data of RTT150 and
1.3-m MDM telescopes during the time interval from 5 to 14 hours after the
burst. Therefore, the power law probably does not provide a good fit to the
data even in this short time interval. Alternatively, the small difference
in the absolute afterglow flux calibration between the telescopes may also
have an effect here. The mean color indices in the first night of
observations, not corrected for the interstellar absorption, are
$V-R_c=0.49\pm0.02$, $B-R_c=1.33\pm0.03$.

\begin{figure}
 \centering
 \smfigurewocapsmall{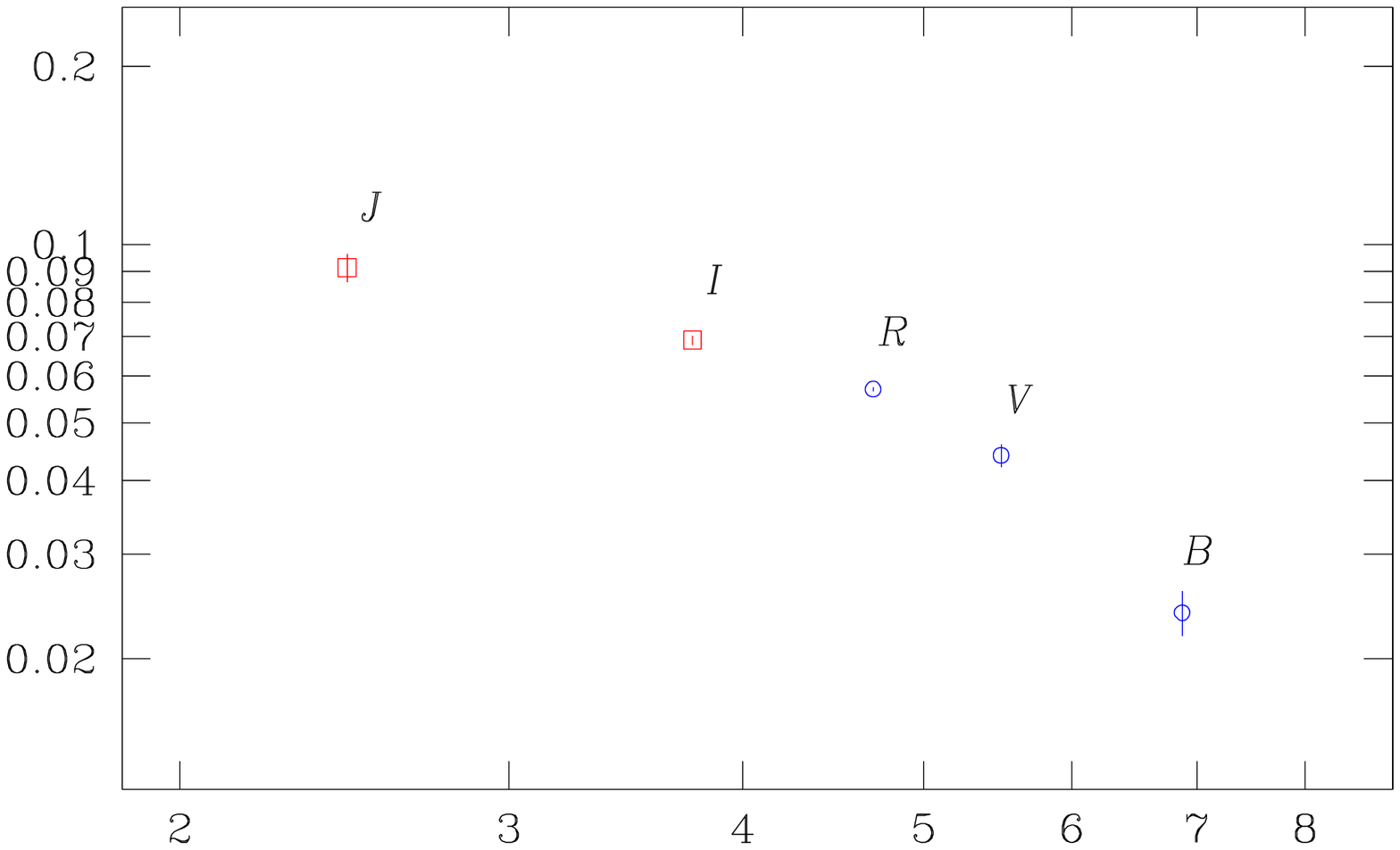}{Frequency, $\times 10^{14}$~Hz}
 {$F_\nu$, mJy}
 \caption{The spectral energy distribution in $B$, $V$, $R$, $I$, $J$ bands
   during 8.7 hours after the gamma-ray burst. The RTT150 data are shown by
   circles; the data in $I$, $J$ bands from \cite{cobb} are shown by
   squares.}
 \label{sed}
\end{figure}

The spectral energy distribution of the afterglow over the 8.7 hours after
the gamma-ray burst, corrected for the Galactic extinction, $E(B-V)=0.066$
\citep{schlegel}, is shown in Fig.~\ref{sed}. The points, corresponding to
$B$ and $V$ bands, are calculated by extrapolating the power low obtained in
the time interval from 5 to 8 hours, assuming the constant colors of the
afterglow at that time. The data in $I$, $J$ bands were taken from the
results of the observations made by 1.3-m CTIO telescope \citep{cobb}.

\begin{figure}
  \centering
  \smfigurewocap{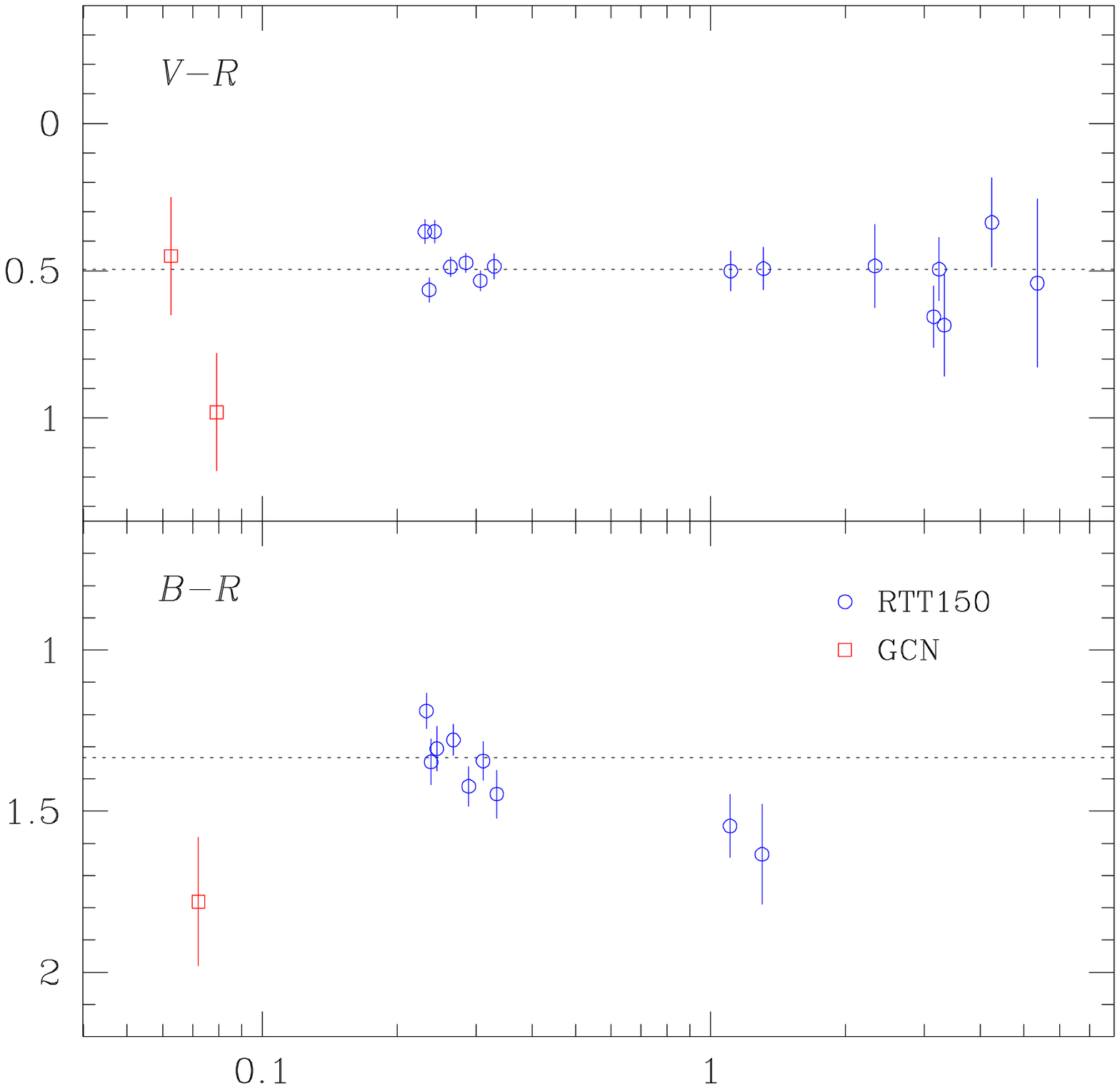}{Time, days}{Color index}
  \caption{The color indices $V\!\!-\!\!R_c$ and $B\!\!-\!\!R_c$ as a
    function of time. The mean values of color indices during the first
    night of observations are shown with dashed lines. The RTT150 data are
    shown by circles, the data taken from GCN Circulars
    \citep{lin,rumyantsev2,brown} are shown by squares.}
  \label{colors}
\end{figure}

The history of afterglow color changes during the all our observations is
shown in Fig.~\ref{colors}. The color index $V-R_c$ proves to be exactly
constant over all our observations. The color index $B-R_c$ changes so that
the afterglow gets somewhat more red in the second night of our observations
with the color $B-R_c=1.59\pm0.09$. Before the time, when our telescope was
pointed to the afterglow, its color can be estimated using the data in $R_c$
band from ground based telescopes: 1-m of Lyulin observatory \citep{lin} and
1.5-m telescope of Maidanak high-altitude observatory \citep{rumyantsev2}
and the data in $BV$ bands from UVOT telescope onboard of SWIFT
\citep{brown}. Corresponding points are shown in the Fig.~\ref{colors} as
squares. Probably, at $t-t_0=1.5$~h, the afterglow was also redder than
during our observations from 5.4 to 9 h after the burst.

The results of measurements of the afterglow light curve in $BVR_c$ bands
are given in Table~\ref{tab:mags}, which contain the values plotted in
Fig.~\ref{lc} and used for color calculation in Fig.~\ref{colors}. In
comparison to the results published in GCN circulars \citep{khamitov}, the
data were reprocessed carefully and the measurements here are more reliable.

\begin{table}
  \centering
  \caption{Photometric measurements~\label{tab:mags}}
  \smallskip
  \begin{tabular}{clcccc}
  \hline
  \hline
  Time, days & Filter & magnitude & error\\
  \hline
0.22679 & \phantom{00}$R_{c}$ & 18.676 & 0.056 \\
0.22892 & \phantom{00}$R_{c}$ & 18.992 & 0.033 \\
0.23067 & \phantom{00}$V$ & 19.302 & 0.041 \\
0.23242 & \phantom{00}$B$ & 20.133 & 0.056 \\
0.23417 & \phantom{00}$R_{c}$ & 18.931 & 0.032 \\
0.23583 & \phantom{00}$V$ & 19.527 & 0.042 \\
0.23813 & \phantom{00}$B$ & 20.321 & 0.072 \\
0.24058 & \phantom{00}$R_{c}$ & 18.988 & 0.042 \\
0.24254 & \phantom{00}$V$ & 19.364 & 0.040 \\
0.24525 & \phantom{00}$B$ & 20.316 & 0.070 \\
\dots & \phantom{00}\dots & \dots & \dots\\
  \hline
  \end{tabular}
  \medskip

  \begin{minipage}{0.9\linewidth}
    \footnotesize \emph{Note}: --- The complete version of this table
    is available in electronic version of the journal, and also at:
    http://hea.iki.rssi.ru/grb/060526/indexeng.html
  \end{minipage}

\end{table}

\section*{DISCUSSION}

In this paper we present the results of detailed observations of the light
curve of GRB 060526 optical afterglow carried out with Russian-Turkish 1.5-m
Telescope (RTT150). Our observations provide about a half of all photometric
measurements obtained for this afterglow. The upper limits to afterglow
variability during the first night of our observations were obtained, and
the history of afterglow color changes is measured in detail.

The afterglow observations of GRB 060526 were also discussed in detail in
the paper by \cite{dai}, who also used our data taken from GCN Circulars
\citep{khamitov}, where it was pointed out that the changes of the power law
index occurred approximately simultaneously in optical and X-ray bands.
This should be expected in the case if this break occur in result of change
of the geometry of the motion of the ejecta. For example, this may occur
when the gamma factor of the jet becomes smaller than the reciprocal of its
opening angle and a sideways expansion of the jet begins.

\begin{figure}
  \centering \smfigurewocapsmall{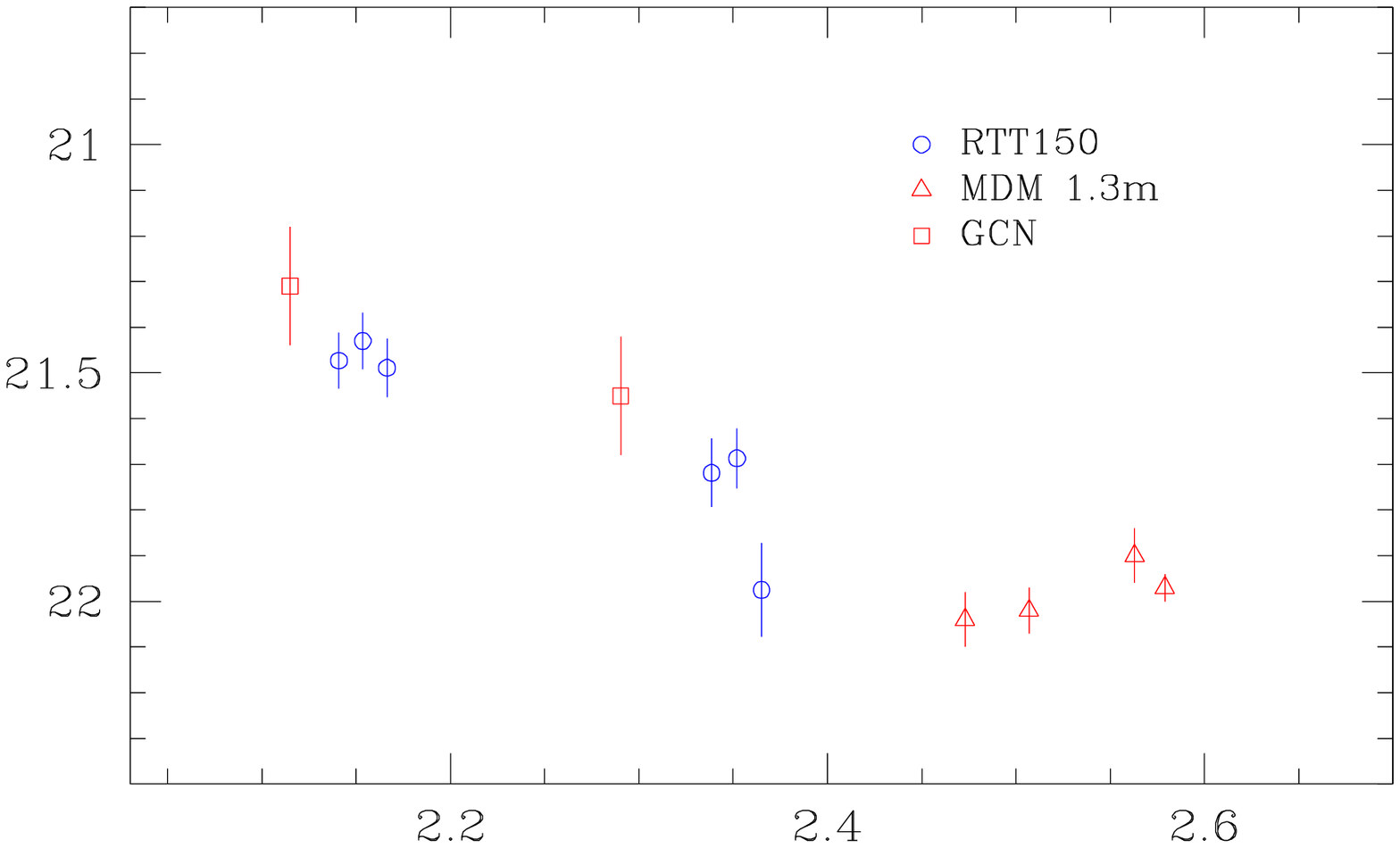}{Time, days}{Magnitude,
    $R_c$}
  \caption{The lightcurve of afterglow of GRB 060526 in \textit{$R_c$} band.
    The RTT150 data are shown by circles, the 1.3-m \textit{MDM} telescope
    data \citep{dai} by triangles, the data taken from GCN Circulares
    \citep{rumyantsev2,kann1} --- by squares.}
  \label{lc2d}
\end{figure}

Dai et al. (2007) also discussed the presence of optical variability in the
afterglow (Fig.~\ref{lc}). They noted the presence of flare activity
starting from $t-t0\sim16$~h and found that the $\delta t/t - \Delta
F_\nu/F_\nu$ relation for these flares poorly satisfies the constraints for
various variability models of ultrarelativistic jet \citep{ioka}. Our data
allow to make even stronger conclusions. Fig.~\ref{lc2d} shows a segment of
the afterglow light curve about two days after the GRB. We see that,
according to our measurements, a sharp change occurs in the light curve at
$t-t_0\approx 2.35$~days --- the brightness declines by $\approx0.3^m$ in
approximately 20 min. This change corresponds to $\delta t/t \approx 0.0055$
and $\Delta F_\nu/F_\nu \approx 0.3$ the constraints on the variability of
the observed emission from an ultrarelativistic jet obtained by \cite{ioka}
are violated with a large margin.

The main condition from which the constraints on the variability were
obtained by \cite{ioka} is an ultra-relativistic motion of the jet towards
the observer. Actually, however, the jet motion may well become moderately
relativistic even on $\sim1$~day time scale at higher densities of the
ambient interstellar medium or in the presence of dense stellar wind
\citep{burenin07}.

If the motion is no longer ultrarelativistic, then fast variability similar
to that discussed above (Fig.~\ref{lc2d}) can be explained much more easily,
for example, by the presence of density inhomogeneities in the stellar wind
around the GRB source. This may be expected, since the stellar winds around
Wolf-Rayet stars are highly inhomogeneous and have a clumpy structure
\citep[e.g.,][]{crowther}. In this case, the sizes of the density
inhomogeneities are $c \delta t \sim 3\cdot10^{13} - 3\cdot10^{14}$~cm
(Figs.~\ref{lc} and \ref{lc2d}). Such inhomogeneities are possible in the
winds from Wolf-Rayet stars at large distances from the star \citep{lepine}.

The power law index of the lightcurve at the end of our observations at
$t-t_0>3$~days, $\alpha\approx -3.4$ (Fig.~\ref{lc}) seems to be too steep
for the non-relativistic stage of shell expansion phase, when the shell
expands according to the Sedov solution for a strong explosion and the
emission is generated by electrons accelerated at the shock front
\citep{huang03}. However, this slope is also too steep for the emission of
an ultra-relativistic jet after the beginning of its sideways expansion
\citep{dai}. Note, that this slope was measured on a very short time scale
and may not reflect the rate of afterglow decline on a longer time scales.
The sharp decrease in radiation flux after the individual flares can be
determined, for example, by the shapes of individual density inhomogeneities
or even by the cooling time of optically emitting electrons.

We also note, that the spectral energy distribution in optical and infrared
bands is not a power law, already at 9 hours after the burst and during at
least a day of observations (Fig.~\ref{sed}, \ref{colors}). This may suggest
that there is a thermal radiation component in the spectrum that can
originate only in a nonrelativistic shock, where the particles at the front
are not accelerated to ultrarelativistic energies. On the other hand, this
may suggest the presence of absorption in a dense medium around the source.

The observed variability can probably be also explained with the assumption
of the ultrarelativistic motion of the jet, for example, assuming the
presence of late inner engine activity \citep{ioka}. Regarding the afterglow
from GRB 060526, this question was discussed in detail by \cite{dai}. The
explanation related to the transition to subrelativistic motion is offered
here only as a possibility. This would require a higher density of the
ambient interstellar meduim or a dense stellar wind around the source, which
are, actually, expected in star-forming regions and around massive
Wolf-Rayet stars \citep[see also][]{burenin07}.

\acknowledgements

We are grateful to anonymous referee for useful comments. This work is
supported by the Russian Fund for Basic Researches (grants 05-02-16540,
07-02-01004), by Russian Government Program of Leading Science Schools
Support (grants 784.2006.2, 1100.2006.2), and by the programs of Russian
Academy of Sciences (P-04, OFN-17). The work of one of the coauthors (RB)
was also supported by grant of President of Russian Federation
MK-4064.2005.2.

\end{document}